\documentclass[compsoc, conference, a4paper, 10pt, times]{IEEEtran}

\usepackage{cite}
\usepackage{amsmath,amssymb,amsfonts}
\usepackage{algorithmic}
\usepackage{textcomp}
\usepackage{xcolor}
\usepackage[hidelinks]{hyperref}

\usepackage{ifthen}
\usepackage{color}
\usepackage{graphicx}
\usepackage{float}
\usepackage{todonotes}
\usepackage{soul}
\usepackage{multirow}

\usepackage{tikz}
\usepackage{tabularx}
\usepackage{booktabs}
\usepackage{wasysym}
\usepackage{forloop}
\usepackage{enumitem}
\usepackage{comment}
\usepackage{ulem}
\usepackage{authblk}

\usepackage[all]{nowidow}

\def\BibTeX{{\rm B\kern-.05em{\sc i\kern-.025em b}\kern-.08em
    T\kern-.1667em\lower.7ex\hbox{E}\kern-.125emX}}

\newif\ifdiff
\difftrue

\newcommand{\Qq}[1]{\textbf{#1}}

\newcounter{qr}

\newcommand{\Qline}[1]{\noindent\rule{#1}{0.6pt}}

\newcounter{ql}

\newenvironment{Qlist}{%

\begin{itemize}[topsep=0.1em]
}{%
\end{itemize}
} 

\newlength{\qt}

\newcounter{itemnummer}
\newcommand{\Qitem}[2][]{
\ifthenelse{\equal{#1}{}}{\stepcounter{itemnummer}}{}
\ifthenelse{\equal{#1}{a}}{\stepcounter{itemnummer}}{}
\begin{enumerate}[topsep=2pt,leftmargin=2.8em]
\item[\textbf{\arabic{itemnummer}#1.}] #2
\end{enumerate}
}

\definecolor{bgodd}{rgb}{0.8,0.8,0.8}
\definecolor{bgeven}{rgb}{0.9,0.9,0.9}
\newcounter{itemoddeven}
\newlength{\gb}
\newcommand{\QItem}[2][]{
\setlength{\gb}{\linewidth}
\addtolength{\gb}{-5.25pt}
\ifthenelse{\equal{\value{itemoddeven}}{0}}{%
\noindent\colorbox{bgeven}{\hskip-3pt\begin{minipage}{\gb}\Qitem[#1]{#2}\end{minipage}}%
\stepcounter{itemoddeven}%
}{%
\noindent\colorbox{bgodd}{\hskip-3pt\begin{minipage}{\gb}\Qitem[#1]{#2}\end{minipage}}%
\setcounter{itemoddeven}{0}%
}
}

\begin{document}

\title{Exploring Smart Commercial Building Occupants' Perceptions and Notification Preferences of Internet of Things Data Collection in the United States}



\author[1,2]{Tu Le}
\author[1]{Alan Wang}
\author[3]{Yaxing Yao}
\author[4]{Yuanyuan Feng}
\author[1]{Arsalan Heydarian}
\author[5]{Norman Sadeh}
\author[1,2]{Yuan Tian}
\affil[1]{University of Virginia}
\affil[2]{University of California, Los Angeles}
\affil[3]{University of Maryland, Baltimore County}
\affil[4]{University of Vermont}
\affil[5]{Carnegie Mellon University}

\maketitle

\begin{abstract}
Data collection through the Internet of Things (IoT) devices, or smart devices, in commercial buildings enables possibilities for increased convenience and energy efficiency. However, such benefits face a large \textit{perceptual} challenge when being implemented in practice, due to the different ways occupants working in the buildings understand and trust in the data collection. The semi-public, pervasive, and multi-modal nature of data collection in smart buildings points to the need to study occupants' understanding of data collection and notification preferences. We conduct an online study with 492 participants in the US who report working in smart commercial buildings regarding: 1) awareness and perception of data collection in smart commercial buildings, 2) privacy notification preferences, and 3) potential factors for privacy notification preferences. We find that around half of the participants are not fully aware of the data collection and use practices of IoT even though they notice the presence of IoT devices and sensors. We also discover many misunderstandings around different data practices. The majority of participants want to be notified of data practices in smart buildings, and they prefer push notifications to passive ones such as websites or physical signs. Surprisingly, mobile app notification, despite being a popular channel for smart homes, is the least preferred method for smart commercial buildings.





\end{abstract}



\section{Introduction}

The Internet of Things (IoT), or smart devices, have increasingly made their way into various physical environments, transitioning them into ``smart environments''. These smart devices have introduced significant benefits to users and society at large. Continuous monitoring of indoor environmental conditions and user behaviors in smart devices-equipped buildings can help reduce energy consumption as well as enhance users' comfort and well-being~\cite{wagner2018exploring, allen2020healthy}. For example, Lu et al.~\cite{lu2010smart} shows that using sensors to intelligently control the home's heating, ventilation, and cooling (HVAC) system can achieve a 28\% energy saving. Figueiro et al. also shows how properly applied light exposures can increase alertness and circadian entertainment \cite{figueiro2020light}.

\textbf{Problem.} Despite the numerous potential benefits of making environments ``smarter'', the transition may also introduce great challenges due to the potential privacy issues~\cite{lipford2022privacy}. Continuous data collection can expose more data than anticipated by the users, and the collected data can be shared with third parties~\cite {ren2019exposure, cruz2023augmented}. One particular privacy issue in these environments relates to occupants' awareness of these smart devices and their data collection and use practices. Research has shown that occupants in smart environments have significant privacy concerns, yet the level of transparency regarding the data practices in smart environments and their ability to control these data practices are limited~\cite{harper2022user}. To increase the transparency of data practices and raise occupants' awareness of data practices in smart environments, research has proposed various mechanisms, such as notifications via mobile devices, network monitoring through web apps, ambient lights, and sounds, etc.~\cite{voit2022exploring, das2018personalized, feng2021design, pappachan2017privacy, jin2022exploring}. However, prior research primarily focuses on smart homes, with less focus on other more public smart environments. Several key differences exist between managing privacy in homes versus commercial buildings, making smart building privacy notification a novel and challenging problem. First, in a home, the same people affected by the potential privacy invasions are mostly capable of changing or removing the offending devices. In contrast, in a commercial building, the occupants might be less aware of the data collection and might feel they are less in control of their privacy. Second, the occupants might have a different mental model when facing the commercial buildings' pervasive data collection compared to their own homes. Finally, smart building data collection is multi-modal, pervasive, and large-scale. The privacy notifications, if not well designed, will cause user apathy or misunderstanding. As a result, there is a need to comprehensively understand users' awareness, perceptions of data collection, and privacy notification preferences in the smart commercial building environment to inform the design of smart building privacy notifications.

\textbf{Research Goal.} In this paper, we focus on smart commercial buildings, an understudied yet important smart environment in the privacy literature. We use ``smart commercial buildings'' to denote commercial buildings that are equipped with smart devices (e.g., Internet-connected security cameras) and sensors (e.g., smart water meters), and use ``occupants'' to denote people who work in or regularly enter these buildings. We aim to understand occupants' awareness of smart devices in these buildings, as well as their preferences in receiving notifications about smart devices and their associated data practices. Our scope focuses on occupants in the US.

\textbf{Importance.} This research is significant for two reasons. First, due to the nature of occupants' tasks and activities in smart commercial buildings, the privacy implications can be different from those in other environments, such as smart homes. Particularly, how to appropriately handle privacy notifications in smart buildings remains an open issue. In addition, existing techniques for maintaining privacy in other IoT environments such as smart homes are unlikely to apply in the smart building context. For example, the power dynamics in smart commercial buildings (e.g., employers vs. employees, administrators vs. tenants) may influence how occupants perceive privacy. 
Second, recent privacy regulations around the world have also mandated the disclosure of certain data collection practices in public places. For example, both the General Data Protection Regulation (GDPR) and the California Consumer Privacy Act (CCPA)
protect users' control over any personal information a business collects about them ~\cite{GDPR, goddard2017eu, CCPA,goldman2020introduction}. As a result, it is important to understand occupants' privacy expectations and study their privacy notification preferences in smart commercial buildings. It is also timely for building owners to understand the mechanisms underlying privacy expectations and build a fiduciary relationship with their occupants.


\textbf{Research Questions.} In this paper, we aim to answer the following research questions:
\begin{itemize}
    \item \textbf{RQ1:} What are the occupants' perceptions of data collection in smart commercial buildings?
    \item \textbf{RQ2:} What are the occupants' notification preferences for data collection in smart commercial buildings in different contexts?
    \item \textbf{RQ3:} What potential factors affect occupants' notification preferences for data collection in smart commercial buildings in different contexts?
\end{itemize}

\textbf{Our Study.} To answer these research questions, we conducted an online study of 492 participants in the US.  
Our participants are people who have worked in smart commercial buildings. We use the term \textit{(smart) commercial building} throughout the paper to indicate our participants' indoor workplaces that deploy IoT devices/sensors other than their homes. In our study, we explored the participants' awareness and perception of data collection in the buildings and used a series of questions to identify whether they want to be notified on the different modalities of data collection, why or why not to be notified, what type of information should be communicated, and how they would like to receive these notifications. We design three hypothetical scenarios based on common IoT devices in smart buildings (i.e., Bluetooth beacons, cameras, smart meters) to ask about participants' privacy notification preferences.

\textbf{Key Findings.} Our results suggest that many participants are unaware of or have misunderstandings of the data collection in smart commercial buildings. Even participants who are highly confident about their knowledge of IoT devices, still have misunderstandings about data collection purposes and data access of smart devices in smart commercial buildings. For example, few people understood that Bluetooth beacon's data would be used for localization even though localization is in fact its primary data purpose. One participant even incorrectly assumed that Bluetooth beacon could get unauthorized access to his/her phone.
In terms of their notification preferences, the majority of our participants (91\%) indicated their willingness to receive notifications about the data practices regardless of their prior knowledge of smart devices. We also found that email was the most desired channel to deliver privacy-related notifications, while some participants preferred other channels (e.g., physical signs) depending on the scenarios. Our data helps us identify several factors that may impact our participants' notification preferences, such as their awareness of data collection and confidence in their knowledge of smart devices.

\textbf{Contributions.} This work contributes to privacy research and human-centered computing in several aspects:
\begin{itemize}
    \item We provide a comprehensive user study to understand occupants' awareness and perceptions of data collection in smart commercial buildings, generating important empirical evidence in this area of research.
    \item Our study provides a systematic understanding of occupants' preferences for privacy notifications in smart commercial buildings and the potential factors that impact their preferences. 
    \item We draw implications for designing a transparent data collection framework for future generations of smart commercial buildings.
\end{itemize}

\textbf{Paper Outline.} The rest of the paper is organized in the following way. We first introduce the related work in Section~\ref{sec:relatedwork} and then explain the design of our user study in Section~\ref{sec:method}. We present our data analysis and results about the user study in Section~\ref{sec:results}. We then discuss the privacy law implications, the suggestions for improving smart building data collection transparency, the limitations of our study, the potential future work in Section~\ref{sec:discussion}, and conclude the study in the end.



\section{Related Work}
\label{sec:relatedwork}
This section discusses the previous work and how our work is different. The related work is presented as three themes: IoT in Smart Buildings, IoT Privacy, and Privacy Notifications for IoT.

\subsection{IoT in Smart Buildings}
While there are many types of sensors deployed in buildings to collect information about the occupant and the surrounding environment, our study mainly focuses on the three types of smart devices that are popular and have been demonstrated to be privacy-invasive, i.e., Bluetooth beacons, cameras, and smart meters. 

For Bluetooth beacons, Caesar et al.~\cite{caesar2020location} demonstrated that Bluetooth technology can be used maliciously to track occupant location. For example, a smartphone can be used to secretly monitor nearby Bluetooth Mesh activity and reference user location through transmissions, or an app installed on smartphones can be used to track a user within a Bluetooth Mesh network. 

For cameras, besides being able to identify users directly, cameras also reveal information that might be difficult or impossible to detect with the naked eye. For example, Davis et al. \cite{davis2014visual} showed that audio signals could be extracted through motion magnification of video data. The same motion magnification technique has also been shown to be able to extract health-related information from users such as blood circulation \cite{rubinstein2014analysis}. 

For smart energy meters, Jazizadeh et al. \cite{jazizadeh2014unsupervised} demonstrated how Non-Intrusive Load Monitoring (NILM), or measuring electricity consumption using an energy meter at the circuit level instead of the appliance level, can still be disaggregated to extract signals of specific appliance use in households and occupant behaviors. Although network communications among IoT devices can be encrypted to ensure privacy, Acar et al.~\cite{acar2020peekaboo} showed that an adversary can exfiltrate sensitive data from the encrypted traffic. Rondon et al.~\cite{Rondon2020PoisonIvyP, rondon2021survey} demonstrated different attacks on different layers of enterprise IoT systems in smart buildings. 

Besides, Babun et al.~\cite{babun2021iotsurvey} provided an analysis of popular IoT platforms in terms of how they handle vulnerabilities and possible solutions for these platforms. Our work focuses on the occupants' perspectives on IoT devices and their data collection in smart buildings.

\subsection{IoT Privacy}
There has been considerable literature investigating the privacy preferences and factors that affect users' privacy decision-making in IoT scenarios. Yao et al.~\cite{yao2019defending, yao2019privacy} conducted co-design studies to identify key factors for designing smart home privacy controls. In a broader context, Naeini et al.~\cite{naeini2017privacy} showed that participants were more comfortable with data collection in public rather than private settings and were more likely to share data for uses that they find beneficial (e.g., find public restrooms). The collection of biometric data is considered less comfortable than environmental data, and the participants wanted to be notified about the data practices of such information being collected. Different from these previous studies, we focus specifically on smart commercial building occupants rather than the general users and consider participants' background knowledge/confidence in IoT technology. We also explore more detailed perceptions regarding notification preferences and how occupants want to be informed of data collection. Via in situ studies, some previous work found users' privacy concerns or misunderstanding of the facial recognition technology~\cite{zhang2021did, zhang2021facial} and fitness tracker~\cite{velykoivanenko2022fitness}. In particular, Zhang et al.~\cite{zhang2021did, zhang2021facial} studied people's notification preferences, including the frequency of notifications. However, they did not explore modality preferences such as emails versus mobile apps because people might not have an email address in the scenarios they considered. Harper et al.~\cite{harper2022user} conducted an online survey of 81 participants to understand their privacy concerns in the smart building context, focusing on environmental data collection. Our work considers a larger pool of participants, more types of smart devices, and occupants' preferences for different notification schemes.

Other work looked into the influence of friends and experts on privacy decisions~\cite{naeini2018influence}. These studies showed that participants were more influenced when their friends denied data collection than when their friends allowed data collection. In contrast, the participants were more influenced when experts allowed collection than when experts denied data collection. However, after being exposed to a set of scenarios in which friends and experts allowed or denied data collection, the participants were less likely to be influenced in subsequent scenarios. Barbosa et al.~\cite{barbosa2019whatif} presented machine learning models to predict personalized privacy preferences in smart homes and identify factors that could change such preferences. 

Several frameworks were proposed to help users to enforce privacy protections on IoT devices. Apthorpe et al.~\cite{apthorpe2018discover} presented a framework to discover the privacy norms in the smart home context. IoTWatch~\cite{babun2021realtimeiot} allows users to specify their privacy preferences at install time and ensure IoT apps' behaviors match the selections. Kratos~\cite{Sikder2020KratosMM} provides smart home users with access control settings that consider multiple users and devices in a shared space. Cejka et al.~\cite{cejka2019privacy} presented potential countermeasures for privacy issues of smart meters. Wu et al.~\cite{wu2021smart} proposed a privacy-preserving framework to support sensor applications such as occupancy detection while ensuring user privacy. In contrast, our work contributes new insights into smart building data collection from occupants' perspectives to support future designs of systems and frameworks.

\subsection{Privacy Notifications for IoT}
Privacy notifications are a type of privacy notice that informs people or users about the data being collected and using practices of a system, product, or service. 
They are often provided by entities responsible for the disclosed data practices (e.g., data collection, sharing, and processing), as increasing requirements by privacy regulations around the world (e.g., GDPR~\cite{GDPR}, CCPA~\cite{CCPA}).
Although privacy policies are the most common type of privacy notice, they are lengthy and difficult to read~\cite{cranor2012necessary, oeldorf2019overwhelming}. Instead, researchers and practitioners have proposed more effective ways to notify people about data privacy practices, such as concise privacy notices~\cite{ebert2021bolder} and privacy nutrition labels~\cite{emami2020ask}. Moreover, Schaub et al. outlined a design space for more effective privacy notices ~\cite{schaub2015design}.

User-facing privacy notifications, in addition to or in lieu of privacy policies, are very common in the digital world. For example, websites have increasingly adopted GDPR-compliant cookie banners, which automatically pop up when websites detect new users. These banners usually contain a concise privacy notification describing how the website uses cookies to track user data and how users can disable some of them~\cite{GDPRcookies}. 
Another common example is the app permission management framework on smartphones. Both iOS and Android platforms send users just-in-time notifications when apps are trying to access certain sensitive permission on smartphones, along with the choice to allow or deny.
The notifications in both examples are delivered to users through primary channels, which is the same platform or device a user interacts with. Huang et al.~\cite{huang2020iotinspector} presented a tool that examines network traffic in a smart home and informs the user of vulnerabilities or tracking services.

However, in IoT-embedded smart buildings, providing people with effective privacy notices is extremely challenging.
First, since smart buildings have countless IoT devices and sensors collecting data (e.g., energy sensors, lighting, temperature, air quality, etc.), it is possible that the number of notifications can overwhelm building occupants and visitors. This may lead to privacy fatigue~\cite{choi2018role} if users receive too many irrelevant privacy notices.
Second, IoT devices and sensors in smart buildings lack traditional user interfaces (e.g., screens), so it is difficult to deliver privacy notifications through the most intuitive primary channels (i.e., IoT devices and sensors themselves). This means privacy notices need to rely on secondary or public channels (e.g., a website, or physical signs), causing an additional barrier for residents or visitors to receive them. Researchers have recently developed a location-based mobile app, IoT Assistant, capable of notifying people about nearby IoT data privacy in public places~\cite{das2018personalized, feng2021design}. However, little research has examined what types of IoT privacy notices people would like to receive and how to receive them.
Therefore, our study aims to understand people's notification preferences for smart building scenarios to inform the design of more effective privacy notices in smart buildings.

\section{Methodology}
\label{sec:method}
\begin{figure}[htbp]
  \centering
  \includegraphics[width=\linewidth]{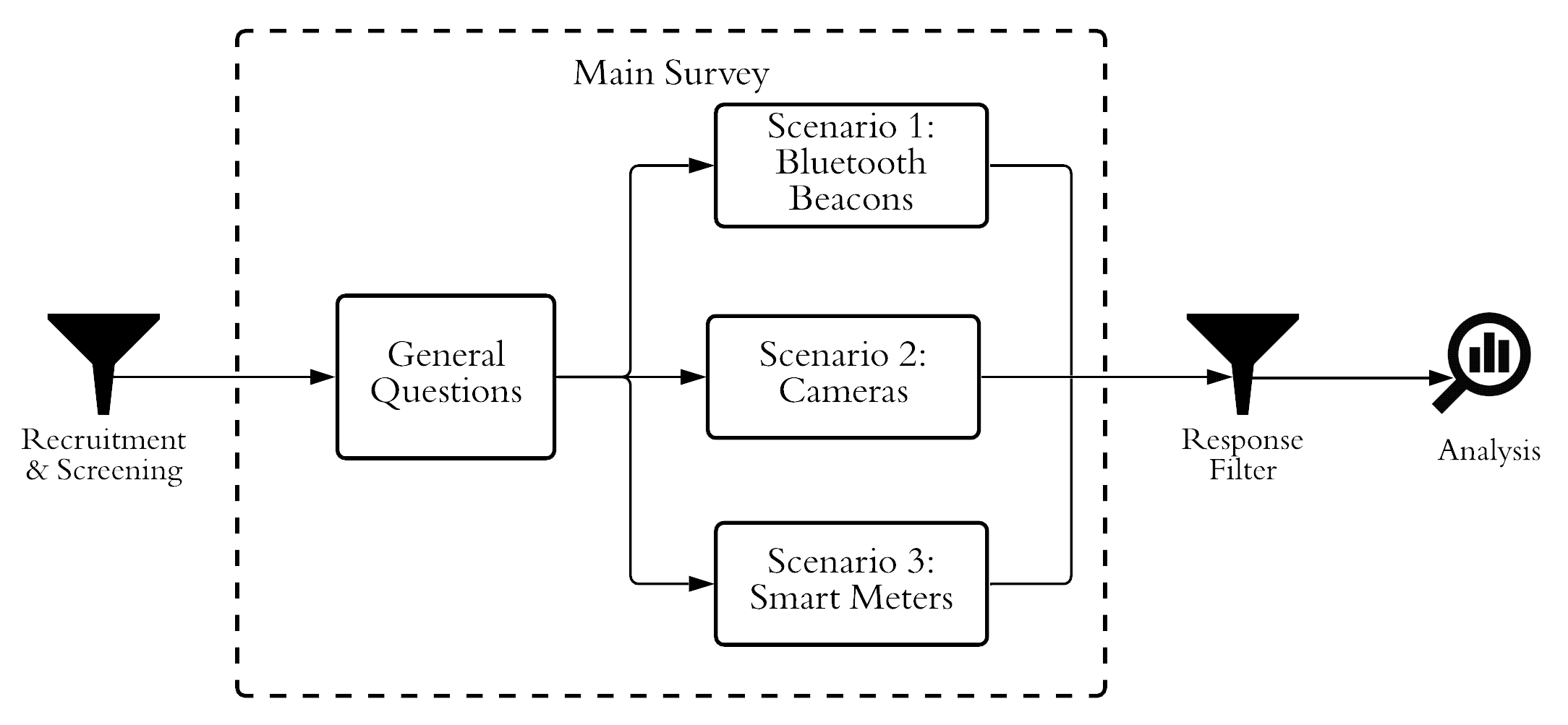}
  \caption{Overview of the study protocol.}
  \label{fig:surveystruct}
\end{figure}

Figure~\ref{fig:surveystruct} shows an overview of our study workflow. We describe our study protocol in detail in this section. In this study, our survey design aims to investigate the awareness and perceptions of data collection (RQ1), the notification preferences (RQ2) of occupants in smart commercial buildings, and the factors that affect occupants' notification preferences (RQ3).

\subsection{Recruitment \& Screening}
Our study was built via Qualtrics and posted on Prolific to recruit participants. To ensure the quality of the responses, only people with at least a 95\% approval rate on Prolific were able to view our study. We also set up Qualtrics to disallow retaking the survey. The participants were required to be adults who are 18 or older, fluent in English, live in the US, and have or are working in a commercial building physically (e.g., offices or retail stores). We further conducted a screening to determine participants' eligibility. If the participants had worked in an indoor workplace other than their home, they were eligible to participate in our study. 

We paid each participant \$0.5 for completing our 1-minute screening and followed up with 597 eligible participants for our main survey. Our main survey (presented in Section~\ref{methodology:mainsurveydesign}) took approximately 9 minutes to complete, and each participant was paid \$1.5 for completing it. The longest completion time was 52 minutes. The completion time included the time it took to read and sign the consent form. Prolific allowed a minimum payment rate of \$8/hour and a recommended rate of \$12/hour. Our payment rate was set to \$12/hour, which matched the recommended amount on Prolific. Note that Prolific may request extra payments based on the completion time. We also asked our participants to leave feedback (if any) for our study and we received no complaints from our participants regarding the payment and survey length.


\subsection{Study Pretest}
Cognitive pretesting and pilot study are two common practices to identify potential issues and biases in surveys, such as priming wording or confusing questions, prior to deployment~\cite{presser2004survey}. We followed an iterative review process in which we repeated the process of running pilot studies to get preliminary results as well as feedback and improving our survey design accordingly until no issues arose. 

We also tested our survey by conducting cognitive interviews with four university students and staff outside of the research team who are from a variety of departments and backgrounds. During the interviews, the participants thought out loud when taking the survey. We noted their thought process and asked them to provide feedback for each survey question (e.g., fixing confusing questions or adding more answer choices). 

As a result, we improved the wording and formatting of survey questions and added additional answer choices to some multiple-choice questions. For example, for questions asking about the context/scenario we give, we highlighted the context/scenario (e.g., ``at your workplace'' or ``in the scenario'') to avoid misunderstanding according to the suggestions from the pilot study. We excluded the pilot study data from our final results to avoid biases.

\subsection{Survey Design}
\label{methodology:mainsurveydesign}
In this section, we describe how we design the survey to answer the research questions.

\subsubsection{Structure and Goals}
The survey includes three sections. 
In the first section, we started with questions to understand participants' general perceptions and preferences in smart buildings. These questions include:
\begin{itemize}
\item Participants' awareness of potential data collection in the smart buildings (answering RQ1);
\item Background questions (e.g., confidence in IoT technology and knowledge of IoT), we analyze whether the answers to the questions influence people's privacy notification preferences (answering RQ3);
\item Pre-scenario questions for occupant's general privacy notification preferences in the smart commercial building setting (answering RQ2). 
\end{itemize}

In the second section, we randomly show the participants one of the three hypothetical scenarios of data collection. The three scenarios include common data collection sensors (i.e., Bluetooth beacons, cameras, and smart meters). We then ask the participants the following two sets of questions:
\begin{itemize}
    \item Questions about their perceptions of data collection in the scenario (answering RQ1);
    \item Post-scenario questions about their privacy notification preferences for the scenario (answering RQ2).
\end{itemize}

Note that we use the same set of questions for the pre-scenario questions and post-scenario questions regarding their privacy notification preferences. The goal is to see whether participants' preferences would change after they are exposed to the scenarios.

In the last section, we ask demographic questions. The answers to these questions are used when we analyze the factors that impact people's privacy notification choices (answering RQ3).

\subsubsection{Detailed Survey Flow}
In the following, we will explain the flow of our survey and how we collect responses from the participants. 

\begin{itemize}
\item Questions about participants' awareness of potential data collection in the smart buildings (answering RQ1) and background (answering RQ3): 
To understand the participants' experience of working in a smart building environment, we first ask them to self-report if they work in a smart building. We present a set of smart devices (e.g., cameras, sensors, smart TV, etc.) and ask the participants to select what devices they notice at their workplace. The participants are able to select multiple devices and have an ``other'' option to enter others. We then ask the participants to report if they are aware of the data collection at their workplace (Yes/No). Note that the answers to these questions are used for two analyses: (1) understanding participants' awareness of data collection (answering RQ1); (2) understanding if awareness of the data collection would impact people's privacy notification choices (answering RQ3).



\item Pre-scenario questions for users' general privacy notification references (answering RQ2): To understand our participants' general notification preferences for data collection in smart buildings, we used a series of questions to identify whether they want to be notified, why/why not, what should be notified, and how they want to be notified. First, we asked them if they want to be notified about the presence of the devices collecting data (Yes/No). We followed up with a free-text question to let them explain their reasons. We then presented a list of information about data collection and asked the participants to select what they want to be notified of. Next, we presented a list of notification means (e.g., physical sign, email, mobile, etc.) and asked the participants to select the means of notification that they prefer. Note that for these multiple-choice questions, the participants can select multiple items and have the option to enter additional text answers. We later repeated this series of notification preferences questions after presenting the scenarios to understand the participants' preferences specifically for each scenario.

\item Scenarios-based questions: To understand users' perception of data collection, and privacy notification preferences, we designed three common data collection scenarios for the participants to check during the survey. Each scenario represents a type and purpose of data collection with different devices typically found in a smart commercial building. We study three devices (Bluetooth beacons, cameras, and smart meters) because they are popular data collection devices in smart buildings and they collect personal data about individuals. Our goal is to study how the perceptions of data collection and notification preferences differ across scenarios. We randomly assigned the participants into three groups. Each group was presented with one of the three scenarios below:
\begin{itemize}
    \item Scenario 1 (Bluetooth beacons): Suppose your employer installs Bluetooth beacons (devices that wirelessly broadcast a unique identifier to nearby electronic devices) at your workplace. These beacons are used to collect location and movement information to understand how the space is used.)
    \item Scenario 2 (Cameras): Suppose your employer installs video surveillance cameras at your workplace that collect photo/video footage to ensure workplace security.
    \item Scenario 3 (Smart meters): Suppose your employer installs smart meters at your workplace that collect data about human activities and resource usage (e.g., energy consumption, bathroom usage) to monitor and optimize the resource consumption.
\end{itemize}

\item Questions about participants' perception of data collections (answering RQ1): After presenting the scenario descriptions, we asked the participants about their perceptions of data access. In particular, they were asked to select from a list of entities that could potentially (e.g., building manager, supervisor, government, etc.) have access to data about them, who they are comfortable with, who they think will benefit from having access to this data, and select/add what purposes they think the data might be used for. We further reused the aforementioned series of notification preferences questions to understand their preferences specifically for the presented scenario.

\item Post-scenario questions about participants' privacy notification preferences (answering RQ2): We repeat the questions about privacy notification preferences in the pre-scenario section to check if users' preferences would be different for specific scenarios.

\item Demographics: Finally, we asked our participants a set of demographic questions, including gender, age, education, and income. 

\end{itemize}

\subsection{Data Analysis}
\label{methodology:analysis}
Our data includes multiple-choice (only 1 choice can be selected), multiple-response (multiple choices can be selected), 5-point Likert scale, and free-text data types. We used Chi-square test (for categorical data) and Kruskal-Wallis H test (for Likert scale data) to quantitatively analyze the responses across 3 scenario groups. We also conducted follow-up Bonferroni post-hoc tests for identifying statistical significance from pair-wise comparisons. 

For multiple-response questions, we coded each item into its variable holding a Yes or No value. Yes value means the participant selected the item and No means otherwise. We then treated these new variables similarly to those of multiple-choice questions. 

For free-text responses, 4 researchers in our group independently coded a subset of the qualitative data. We first developed and agreed on a code book to capture the themes. We then used this codebook to code the data independently. Each entry in the dataset was coded by 2 researchers. After finishing independent coding, we discussed the codes as a group to resolve conflicts and finalize the codes. 

\subsection{Ethical Considerations}
We worked closely with our Institutional Review Board (IRB) and iteratively updated our study protocol. Our protocol did not receive any obligations or constraints from the IRB. Before the study, we asked the participants to read our consent form carefully and sign it to participate in the study. Participation in our study was voluntary and anonymous. When collecting the data, we did not collect any personally identifiable information except for their Prolific ID (a randomly-generated string of numbers and letters) for payment purposes. All data was securely stored and could only be accessed by the research team. Our survey instrument is attached in Appendix~\ref{appendix:surveyinstrument}.

\section{Results}
\label{sec:results}

This section presents our findings from the user study and how they answer our research questions.

\subsection{Overview}
Our study contributes to a new understanding of people's awareness, perceptions, and notification preferences of IoT devices and the associated data collection behaviors in commercial smart buildings. In general, we found that about half of the participants reported being aware of the data collection by IoT devices at their workplaces while the other half were unaware. However, we observed variances among participants' perceptions regarding who may access their data, whether they are comfortable with their data being accessed, and who might benefit from their data. 

Furthermore, we also unpacked our participants' preferences on whether they would like to receive notification about their data collection, and if so, what information they would like to know and how they want to be notified. Our results highlight the need for notifications as over 90\% of our participants want to be notified. However, they have different expectations of the notification content and modality, suggesting the need for context- and device-dependent notification mechanisms. 

Our study also suggested a few factors that could influence participants' preferences of what and how notifications should be delivered. For example, participants who were aware of the IoT devices' data collection would prefer to know how their collected data is used.

In the remainder of this section, we first summarize the demographics and backgrounds of our study participants, then we present our detailed findings based on the three research questions we listed.

\subsection{Participants}

\textbf{Screening and validation:} We received 800 responses from our screening. 597 participants qualified for our screening and received our invitation to the study. Eventually, we received 564 responses for our main survey. We filtered out invalid responses such as incomplete responses (including meaningless ones that the participant entered only white spaces into all free-text answer boxes), and responses that failed our attention checks. As a result, we removed 8 invalid responses from our dataset. These removed responses include 6 duplicates and 2 attention-check fails.

\textbf{Considering participants whose workplace has IoT devices:} As our study focuses on occupants in smart commercial buildings, we asked the participants to report what devices they noticed at their workplace. We included the ``None'' answer to filter out participants who had no experience with IoT devices at all. We excluded 64 such responses from our dataset. Our final dataset includes responses from 492 participants. 

\begin{table}[htbp]
\begin{center}
\begin{tabular}{|l|l|l|}
\hline
                      & \textbf{Responses} & \textbf{Percentage} \\ \hline
\multicolumn{3}{|l|}{\textbf{Gender}}                            \\ \hline
Male                  & 276                & 56.1\%              \\ \hline
Female                & 208                & 42.3\%              \\ \hline
Non-binary            & 7                  & 1.4\%               \\ \hline
Prefer not to answer  & 1                  & \textless{}1\%      \\ \hline
\multicolumn{3}{|l|}{\textbf{Age}}                               \\ \hline
18 - 24               & 95                 & 19.3\%              \\ \hline
25 - 34               & 226                & 45.9\%              \\ \hline
35 - 44               & 118                & 24\%                \\ \hline
45 - 54               & 35                 & 7.1\%               \\ \hline
55 - 64               & 14                 & 2.8\%               \\ \hline
65 and above          & 4                  & \textless{}1\%      \\ \hline
\multicolumn{3}{|l|}{\textbf{Education}}                         \\ \hline
Some high school      & 2                  & \textless{}1\%      \\ \hline
High school graduate  & 36                 & 7.3\%               \\ \hline
Some college          & 87                 & 17.7\%              \\ \hline
Associate's degree    & 40                 & 8.1\%               \\ \hline
Bachelor's degree     & 207                & 42.1\%              \\ \hline
Graduate degree       & 117                & 23.8\%              \\ \hline
Prefer not to answer  & 3                  & \textless{}1\%      \\ \hline
\multicolumn{3}{|l|}{\textbf{Annual personal income}}            \\ \hline
Less than \$10,000    & 42                 & 8.5\%               \\ \hline
$10,000 - $24,999     & 75                 & 15.2\%              \\ \hline
$25,000 - $49,999     & 125                & 25.4\%              \\ \hline
$50,000 - $74,999     & 103                & 20.9\%              \\ \hline
$75,000 - $99,999     & 55                 & 11.2\%              \\ \hline
$100,000 - $149,999   & 48                 & 9.8\%               \\ \hline
\$150,000 and greater & 33                 & 6.7\%               \\ \hline
Prefer not to answer  & 11                 & 2.2\%               \\ \hline
\end{tabular}
\end{center}
\caption{Demographic information (gender, age, education, and annual personal income before taxes) of the participants in our sample.}
\label{table:demographic1}
\end{table}

\textbf{Demographic background:} Among the 492 participants, 56.1\% are male, 42.3\% are female, and less than 2\% non-binary. Our participants skewed towards young (45.9\% are between 25 and 34), highly educated (42.1\% have Bachelor's degree) people in the middle socio-economic class (46.3\% have income between \$25,000 and \$74,999). Table~\ref{table:demographic1} presents the descriptive statistics about the demographic background of our participants.

Regarding the background of IoT technologies in general, over 60\% of our participants indicated a good level of confidence (i.e., above a rating of 3) with the IoT technologies (e.g., voice assistants, smart security cameras, smartwatches, virtual reality, etc.). Most participants reported that they have an average (34.96\%) and above-average (40.04\%) understanding of IoT technologies. 

In Appendix~\ref{appendix:participants}, we present additional details about participants' level of confidence with the IoT technologies and participants' understanding of the IoT technologies (Figure~\ref{fig:confidenceiot} and  Figure~\ref{fig:understandingiot}).




\subsection{Perceptions of Data Collection in Smart Commercial Building (RQ1)}
Our goal is to understand occupants' perceptions of data collection in smart commercial buildings, identifying potential gaps between occupants' perceptions and the IoT data collection transparency. Our data suggested that the participants reported different levels of understanding in terms of the types of data collection at their workplaces. There is a discrepancy between what people think about IoT data collection and what it actually is.

\subsubsection{Awareness of data collection} 
\begin{figure}[htbp]
  \centering
   \includegraphics[width=\linewidth]{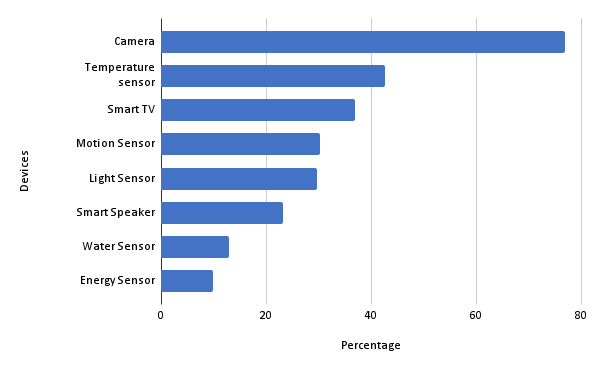}
  \caption{Participants' responses to what devices they notice at their workplace.}
  \label{fig:devicesnotice}
\end{figure}

First, we want to understand if the occupants know of the data collection happening and what kinds of IoT devices they noticed at their workplace. 52.44\% of participants reported being aware of the data collection at their workplace while the other 47.56\% reported being unaware. We asked what devices they noticed at their workplace. Cameras were the most popular device (selected by 76.83\% of participants) that they noticed at their workplace. Among different sensors, most of our participants reported noticing temperature sensors, while the energy sensor was the least noticed one. Figure~\ref{fig:devicesnotice} shows the percentage of responses for what devices the participants noticed at their workplace.

\noindent\fbox{%
  \parbox{\linewidth}{%
    \textbf{Takeaway 1}: Even though people notice IoT devices at their workplace, they are not aware of these devices collecting data. Cameras are the most popular devices that people notice at their workplaces. This indicates that the camera's presences are very obvious to many occupants in a smart commercial building setting. We later present findings about occupants' perceptions and notification preferences of the camera's data collection scenario as compared to two other less obvious scenarios (i.e., Bluetooth beacon and smart meter).
  }%
}


\subsubsection{Perceptions of data access}
\label{result:dataaccess}
We further unpack people's perceptions of data collection from the following three perspectives: (1) who do you think will have access to your data, (2) who are you comfortable with having access to your data, and (3) who do you think will benefit from having access to your data. We provided a list of entities relevant to data collection in smart buildings, which included my building manager, my supervisor, the government, the manufacturer of the devices, my company, and myself. We used Chi-square test to identify the significant associations between the scenario and the selection of each entity. We also conducted a Bonferroni post-hoc analysis to identify the specific pairs of scenarios that have a significant difference.

\textbf{First, we find that the majority of participants (94.5\%) thought that their company would have access to data about them in all 3 scenarios. A few participants (11\%) thought that they [themselves] would have access to their own data.} We identify statistical significance for ``My supervisor'' ($p = 0.002$) and ``The manufacturer of the devices'' ($p = 0.000$) across 3 scenarios. Specifically, for the ``My supervisor'' selection, our pair-wise comparison test shows that in the Bluetooth beacon scenario, significantly more participants thought that their supervisor would have access to their data compared to the smart meter scenario ($p = 0.001$). For the ``The manufacturer of the devices'' selection, we find a significant difference between the camera scenario and each of the other two scenarios, i.e., significantly fewer participants in the camera scenario thought that the manufacturer of the devices would have access to the data as compared to the other 2 scenarios ($p = 0.000$). 

\textbf{Second, we find that most participants felt comfortable with themselves (62.4\%) and their company (53.5\%) having access to their data.} For the Bluetooth beacon and smart meter scenarios, slightly more participants were comfortable with themselves than with their company having access to the data. For the camera scenario, the numbers are equal. We find statistical significance for ``My building manager'' ($p = 0.005$), ``My supervisor'' ($p = 0.027$), and ``The manufacturer of the devices'' ($p = 0.000$) across the 3 scenarios. For ``My building manager'', our pair-wise comparison test shows that significantly fewer participants in the Bluetooth beacon scenario felt comfortable with the building manager having access to their data compared to the smart meter scenario ($p = 0.003$). For the ``My supervisor'' selection, significantly more participants in the camera scenario were comfortable with their supervisor having access to their data compared to the smart meter scenario ($p = 0.022$). For ``the manufacturer of the devices'' selection, significantly more participants in the smart meter scenario were comfortable with the manufacturer of the devices having access to their data compared to the camera ($p = 0.000$) and Bluetooth beacon ($p = 0.001$). 

\textbf{Lastly, when asked about who they thought would benefit from having access to their data, most participants thought that their company would. Only 15.4\% of participants thought that they would benefit from having access to their own data.} We find statistical significance for ``My building manager'' ($p = 0.002$), ``My supervisor'' ($p = 0.000$), and ``The manufacturer of the devices'' ($p = 0.002$). Our pair-wise comparison test indicates the significant difference between the smart meter and each of the other two scenarios regarding ``My building manager'' (camera: $p = 0.003$, Bluetooth beacon: $p = 0.012$) and ``My supervisor'' ($p = 0.001$) selections. Noticeably more people in the smart meter scenario thought that the building manager would benefit from having access to the data while the percentages were similar between the other two scenarios. In contrast, significantly fewer people in the smart meter scenario thought that their supervisor would benefit from having access to the data as compared to the other two scenarios. For the ``the manufacturer of the devices'' selection, we find a significant difference between the camera and each of the other two scenarios (smart meter: $p = 0.004$, Bluetooth beacon: $p = 0.013$). Significantly fewer people in the camera scenario thought that the manufacturer of the devices would benefit from having access to the data. 

\noindent\fbox{%
  \parbox{\linewidth}{%
    \textbf{Takeaway 2}: In a smart commercial building setting, occupants feel more comfortable with having access to their own data over other entities. However, most people do not think that they have access to their own data, yet they would benefit from having access to their own data. Besides, depending on the type of devices that collect data, they have significantly different perceptions about what data are collected and who will benefit from such information. Building managers, supervisors, and the manufacturers of the devices are the three entities that have significant differences in terms of occupants' perceptions across the 3 scenarios (i.e., Bluetooth beacon, camera, and smart meter).
  }%
}

\subsubsection{Perceived purposes of data collection} 
We then asked the participants to select the purposes for that they think the collected data about them could be used. In general, for all 3 scenarios, ``enforcing policies'' is the most selected purpose for data collection (over 75\% of participants). This is a surprising result since most of the time, enforcing policies is not what the collected data is primarily used for. 

Specifically, in the Bluetooth beacon scenario, ``localization'' is the least selected purpose, which is also an interesting result because it is actually the primary usage of Bluetooth beacons. Instead, most participants thought ``enforcing policies'' (76.5\%) and ``user profiling'' (71.6\%) were the purposes of the Bluetooth beacon's data collection. One participant was specifically concerned that Bluetooth beacons could get unauthorized access to his/her phone: ``I would assume the Bluetooth beacon might also be able to access my phone without me realizing (P272).'' 

\begin{figure}[htbp]
    \centering
    \includegraphics[width=\linewidth]{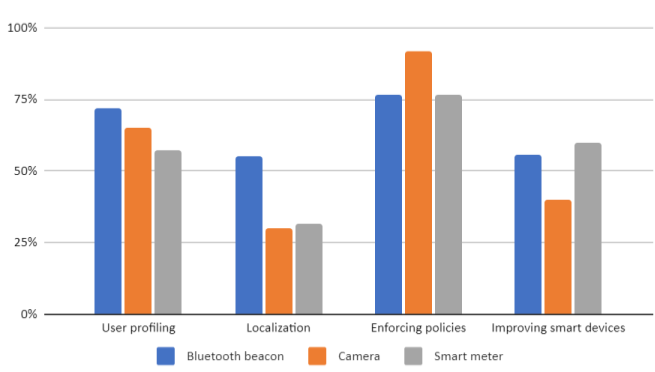}
    \caption{Participants' responses to ``What purposes do you think data about you might be used for?'' across 3 scenarios. Enforcing policies is the most popular selection for all 3 scenarios, which indicates the participants might have misconceptions about the purposes of the devices.}
    \label{fig:data_purposes}
\end{figure}

These results indicate people's misconceptions and lack of knowledge regarding the purposes of data collection in various contexts. Figure~\ref{fig:data_purposes} shows the percentage of participants' responses for what purposes they think the collected data about them might be used.

\noindent\fbox{%
  \parbox{\linewidth}{%
    \textbf{Takeaway 3}: Interestingly, many participants may have misunderstandings about the purposes of the data collection. 76.5\% of the participants in the Bluetooth beacon scenario thought that its data collection is used for enforcing policies. Even though some participants indicated that their understanding of IoT technologies, in general, is at or above average, many of them still believe that the data collection in the three scenarios is for enforcing policy purposes instead of for the functionalities of IoT devices. 
    This may indicate one of the following: 1) participants are not fully aware of the purpose of different IoT devices and they have a misunderstanding of how IoT devices work, or 2) they do not believe that the IoT devices are utilized for the same purposes as they are intended and designed. Thus, it is important to inform people about data collection purposes and data access when designing notifications.
  }%
}

\subsection{Notification Preferences for Data Collection in Smart Commercial Building (RQ2)}



To understand participants' preferences for notifications in smart commercial buildings, we used the following questions regarding the presence of data collection:
\begin{itemize}
    \item Do you want to be notified? Why and why not?
    \item What do you want to be notified about?
    \item How do you want to be notified?
\end{itemize}
We first asked these three questions at the beginning of the survey to get a general overview of the participants' notification preferences. We then presented the hypothetical scenarios and asked these questions again to investigate the participants' notification preferences specifically for the given scenario. As the participants were randomly presented with one of the three scenarios, we need to ensure that the participants in the three groups have a similar level of understanding of IoT technologies so that we can make a comparison among these three groups.
We did not find statistically significant differences among the participants in the three groups, indicating that participants from all three groups have a similar level of IoT knowledge.

\subsubsection{Participants' Notification Preferences}
Next, we present participants' responses to our questions regarding whether they want to be notified about the data collection and what information they want to know in general and in 3 scenarios. Table~\ref{tab:qualitativecate} lists all themes from the participants' responses to why they want or do not want to be notified about the data collection in each scenario.

\textbf{General context.} When asked whether they want to be notified about the presence of the devices collecting data about them at their workplace, the majority of participants (90.65\%) wanted to be notified and only 9.35\% did not want to be notified. 

The majority of participants wanted to be aware of the data collection. Privacy rights and the safety of the data were also considered important to the participants. A participant mentioned ``I want to know how my information may be handled and if it will affect me in my personal life or my work life (P1).'' Other participants expressed privacy concerns such as ``So I don't get spied on without knowing. And being aware of how my facial data is processed by my employer (P9).'' A participant thought that he/she was already being spied on: ``I already know they are spying on us but I would at least like to know from where (P153).''

The few participants that did not want notifications thought that the data collection would not affect them negatively or trusted whoever had access to the collected data: ``I don't think it would affect my performance (P185).'', ``Doesn't bother me much (P151).'', ``Information collected by the company is strictly official and has little or nothing to do with my private life. This information also rests in credible hands (P195).'' Some others were confident that they already knew how things work and that notifications are unnecessary: ``I really understand how it works, I don't need any notification on them (P138).'', ``I personally don't feel it's necessary and also me not knowing would allow me to work and act normally (P281).''

Our data suggest 9 primary reasons why the participants wanted to be notified and 13 reasons why they did not want to be notified about the data collection. For example, increasing awareness of data collection by nearby smart devices remains the top reason why participants would like to be notified (n=246). Notably, 68 participants believed that they have privacy rights in the workplace; as a result, they should be notified of any nearby data collection: ``No matter what that is my right to be notified'' (P172), ``It is one of the rights I have as an employee'' (P178), or ``I believe it is unethical for a company to record information about employees without first telling them about and what information is collected (P196).''


\begin{table*}[!h]
\begin{center}%
\resizebox{\linewidth}{!}{%
\begin{tabular}{|l|l|llll|}
\hline
\multirow{2}{*}{Question}          & \multirow{2}{*}{Categories}           & \multicolumn{4}{c|}{Number of responses}                                                                         \\ \cline{3-6} 
                                   &                                       & \multicolumn{1}{l|}{General} & \multicolumn{1}{l|}{Bluetooth beacon} & \multicolumn{1}{l|}{Camera} & Smart meter \\ \hline
\multirow{9}{*}{Why notified}      & Awareness of data collection          & \multicolumn{1}{l|}{248}     & \multicolumn{1}{l|}{73}               & \multicolumn{1}{l|}{84}     & 79          \\ \cline{2-6} 
                                   & Privacy rights/Ethics                 & \multicolumn{1}{l|}{71}      & \multicolumn{1}{l|}{23}               & \multicolumn{1}{l|}{31}     & 20          \\ \cline{2-6} 
                                   & General concerns/curiosity            & \multicolumn{1}{l|}{40}      & \multicolumn{1}{l|}{29}               & \multicolumn{1}{l|}{21}     & 31          \\ \cline{2-6} 
                                   & Privacy violation                     & \multicolumn{1}{l|}{38}      & \multicolumn{1}{l|}{16}               & \multicolumn{1}{l|}{13}     & 7           \\ \cline{2-6} 
                                   & Pay attention to their behavior       & \multicolumn{1}{l|}{36}      & \multicolumn{1}{l|}{13}               & \multicolumn{1}{l|}{24}     & 12          \\ \cline{2-6} 
                                   & Ensure safety of their data           & \multicolumn{1}{l|}{24}      & \multicolumn{1}{l|}{10}               & \multicolumn{1}{l|}{11}     & 4           \\ \cline{2-6} 
                                   & Trust/Communication/Interaction       & \multicolumn{1}{l|}{18}      & \multicolumn{1}{l|}{20}               & \multicolumn{1}{l|}{18}     & 8           \\ \cline{2-6} 
                                   & Ownership of their own data           & \multicolumn{1}{l|}{12}      & \multicolumn{1}{l|}{18}               & \multicolumn{1}{l|}{14}     & 23          \\ \cline{2-6} 
                                   & Understand the associated benefits    & \multicolumn{1}{l|}{4}       & \multicolumn{1}{l|}{5}                & \multicolumn{1}{l|}{2}      & 1           \\ \hline
\multirow{14}{*}{Why not notified} & No violation/negative effects         & \multicolumn{1}{l|}{7}       & \multicolumn{1}{l|}{1}                & \multicolumn{1}{l|}{0}      & 1           \\ \cline{2-6} 
                                   & Nothing to hide                       & \multicolumn{1}{l|}{5}       & \multicolumn{1}{l|}{2}                & \multicolumn{1}{l|}{1}      & 1           \\ \cline{2-6} 
                                   & Understand how it works               & \multicolumn{1}{l|}{4}       & \multicolumn{1}{l|}{2}                & \multicolumn{1}{l|}{1}      & 0           \\ \cline{2-6} 
                                   & Unnecessary/Not useful                & \multicolumn{1}{l|}{4}       & \multicolumn{1}{l|}{2}                & \multicolumn{1}{l|}{4}      & 10          \\ \cline{2-6} 
                                   & Burdensome/It will worry me more      & \multicolumn{1}{l|}{4}       & \multicolumn{1}{l|}{1}                & \multicolumn{1}{l|}{0}      & 3           \\ \cline{2-6} 
                                   & Feel comfortable with devices around  & \multicolumn{1}{l|}{3}       & \multicolumn{1}{l|}{0}                & \multicolumn{1}{l|}{0}      & 1           \\ \cline{2-6} 
                                   & Just don't want                       & \multicolumn{1}{l|}{3}       & \multicolumn{1}{l|}{0}                & \multicolumn{1}{l|}{2}      & 7           \\ \cline{2-6} 
                                   & Not sensitive                         & \multicolumn{1}{l|}{2}       & \multicolumn{1}{l|}{1}                & \multicolumn{1}{l|}{0}      & 2           \\ \cline{2-6} 
                                   & It makes no difference in my behavior & \multicolumn{1}{l|}{2}       & \multicolumn{1}{l|}{1}                & \multicolumn{1}{l|}{0}      & 0           \\ \cline{2-6} 
                                   & No control anyways                    & \multicolumn{1}{l|}{2}       & \multicolumn{1}{l|}{1}                & \multicolumn{1}{l|}{0}      & 0           \\ \cline{2-6} 
                                   & Employer's privilege                 & \multicolumn{1}{l|}{2}       & \multicolumn{1}{l|}{0}                & \multicolumn{1}{l|}{0}      & 1           \\ \cline{2-6} 
                                   & Won't get enough info                  & \multicolumn{1}{l|}{1}       & \multicolumn{1}{l|}{0}                & \multicolumn{1}{l|}{0}      & 0           \\ \cline{2-6} 
                                   & Not related to privacy                & \multicolumn{1}{l|}{1}       & \multicolumn{1}{l|}{1}                & \multicolumn{1}{l|}{1}      & 5           \\ \cline{2-6} 
                                   & Private property                      & \multicolumn{1}{l|}{0}       & \multicolumn{1}{l|}{1}                & \multicolumn{1}{l|}{0}      & 0           \\ \hline
\end{tabular}%
}
\end{center}
\caption{Categories of qualitative responses regarding reasons why the participants wanted or did not want to be notified about data collection}
\label{tab:qualitativecate}
\end{table*}

For what people want to be notified about, we asked the participants to select from a list of different options before they saw the scenarios as well as after reading the scenarios: the presence of data collection, purposes for which your data can be used, how your data can be used, for how long your data can be retained, and who can access your data. The majority of the participants (more than 89\%) wanted to be notified about all of the listed options. Fewer participants (78\%) wanted to know for how long their data can be retained. 
There were no significant differences in participants' preferences on what they would like to be notified about across a general context and 3 scenarios.


\textbf{Scenario 1: Bluetooth beacon.} 91.36\% of our participants wanted to be notified about the presence of the devices collecting data while 8.64\% did not. Most people wanted to be notified of the purposes for which their data can be used. Participants wanted to be aware of any sensitive info that could be inferred from the data: ``That's really creepy and I don’t want to be constantly tracked (P81).'', ``I don't want potentially embarrassing info collected, like how often I use the bathroom (P261).'' Some participants mentioned that the data collection could affect their job: ``The Bluetooth beacon collects some personal data about my performance by interacting with other devices (P198).'', ``So I can decide whether or not it's a dealbreaker for me in keeping the job (P280).''

The majority of people thought that data collected from Bluetooth beacons would be used for enforcing policies and user profiling. A few participants thought that the data could be used for micromanagement, manipulation, profit from selling to third parties, or improving workspace efficiency. One participant mentioned ``I want to know if my actions are being monitored in some way and have the potential to be used against me (P1).''

When asked about who would have access to the collected data, most people thought that their company would have access to the collected data (95.7\%) and that their company would benefit from having the data access (87.7\%). However, the majority of people (67.9\%) were comfortable with themselves rather than their company having access to the collected data. A participant was worried that someone else might know their private activities: ``If my whereabouts are being tracked, I want to know. It also prevents people from being blindsided when they're confronted with information that they thought no one knew about because they were alone when they were doing it (going from A to B, spending too much time somewhere, etc.) (P73).''

\textbf{Scenario 2: Camera.} 92.77\% of our participants wanted to be notified about the presence of the devices collecting data while 7.23\% did not. More people were interested in getting notified about who can access their data and the purposes for which their data can be used. In contrast to the other 2 scenarios, the participants were more concerned about their behaviors being monitored. A few participants did not want their activities to be watched by someone else: ``So I don't do something embarrassing while I think I am alone and no one but me will ever see or know (P429).''

The majority of participants thought that camera data would be used for enforcing policies. One participant said: ``Because if I do any mistake I will correct that (P45).'' Many participants also selected user profiling as the purpose of using camera data. Some participants expressed concerns about past experience with camera data collection: ``I have been through abusive periods of my life revolving heavily around cameras (P50).''

The majority of participants (95.2\%) thought that their company would have access to the collected data. Many people also thought their supervisor and building manager would have access. Some mentioned the IT department and the software provider of the camera. Most participants were comfortable with themselves and their company having access to the collected camera data. Surprisingly, more participants were comfortable with their company than themselves having access (56.6\% vs 55.4\%). Some mentioned that they would like to get help with loss prevention or in case of theft occurs.

Although the participants would like to get help with loss prevention from their company, a few thought that they themselves would benefit from having access to the collected data (16.9\%). Significantly more participants thought that their company would benefit from having access to the data (86.1\%).

\textbf{Scenario 3: Smart meter.} 81.71\% of our participants wanted to be notified about the presence of the devices collecting data while 18.29\% did not. More people were interested in getting notified about who can access their data and the purposes for which their data can be used. 

In contrast to the other 2 scenarios, more participants worried about the ownership of their data. Most said that it is their right to know about their own data being collected: ``Because it's my data and it should belong to me (P241).'' or ``I feel it is my right to have access to the information that is gathered about me (P109).'' Most people who did not want to be notified thought that it is not necessary or not useful: ``I feel it's of no need to enable me to know the presence. I feel it's right to let it do its work without myself having to feel its presence (P11).'' Some participants thought the data collection was not related to privacy: ``Data based on resource consumption is not really privacy related (P7).'' 

Surprisingly, enforcing policies is the purpose of using the collected data that most participants thought. Noticeably fewer people selected ``improving smart devices'' which is actually the main purpose of smart meter data collection. A few people mentioned tracking resource usage and intimidation as the purposes of using smart meter data. Some participants expressed concerns about being monitored: ``It feels like they are getting very close to crossing that bridge of going too far for me at least. I'd like to know what they are tracking/watching if I am an employee (P36).''

In this scenario, most participants (92.7\%) thought that their company would have access to the collected data. A few participants (13.4\%) thought that they themselves would have access. When asked about who they are comfortable with having the access, more participants selected ``Myself'' than ``My company'' (64.0\% vs 45.7\%) which indicates that people preferred to have access to their data besides the company. However, the majority of participants (86.0\%) still thought that their company would benefit from it. One participant added that whoever bought the collected data would benefit from it.

\noindent\fbox{%
  \parbox{\linewidth}{%
    \textbf{Takeaway 4}: In general, most people (90.65\% of participants) preferred to be notified about the presence of the devices collecting data about them at their workplace. Across 3 scenarios, participants' desire for notification is consistent in Bluetooth beacon and camera scenarios. However, about 10\% fewer participants wanted to be notified in the smart meter scenario, suggesting that fewer people are concerned about smart meter data collection. For all 3 scenarios, participants expressed their preference for notification of all information about the data collection activity that we listed. Noticeably, participants in the Camera scenario were more interested in knowing whether their behaviors were being tracked, while those in the Smart meter scenario were more interested in the ownership of the collected data. The results indicate the importance of transparency in data collection.
  }%
}


\subsubsection{Participants' Notification Methods Choices}
\begin{figure}[!h]
  \centering
   \includegraphics[width=\linewidth]{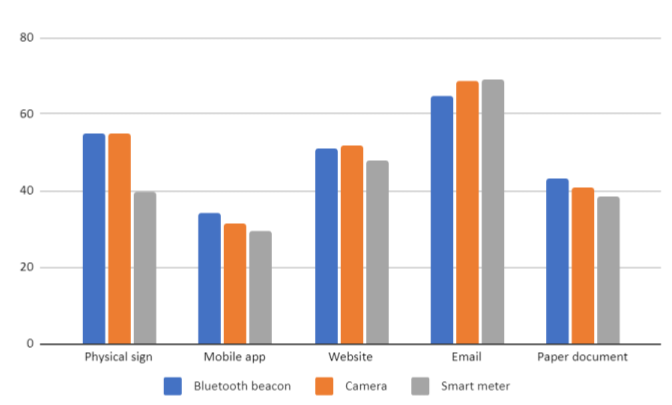}
  \caption{Participants' preferences for how they wanted to be notified across 3 scenarios.}
  \label{fig:notification_how_scenarios}
\end{figure}

Next, we present the participants' preferences of notification methods for data collection in general and in 3 scenarios. Figure~\ref{fig:notification_how_scenarios} show the percentage of what the participants wanted to be notified about and how they wanted to be notified, respectively for 3 scenario groups.

\textbf{Email is the most popular choice across all scenarios and in general.} 69\% of participants selected email as their preferred notification method. Notably, the amount of participants who preferred Email method is significantly larger than other methods. One participant says ``I want to be notified of how my data is being gathered via email on a weekly basis (P113).''

\textbf{Mobile App is surprisingly the least selected choice. } Only 33\% of participants preferred to be notified via mobile app. This choice has a significantly lower number of participants as compared to email (69\%), physical sign (51\%), and website (49\%). Although mobile app has been a popular method of notification~\cite{thakkar2022would}, our finding shows that participants do not prefer to receive mobile app notifications for smart commercial building context.

\textbf{Physical Sign is more preferred in the Bluetooth beacon scenario and the Camera scenario.} Our statistical test shows a significant difference for the ``physical sign'' option between the smart meter scenario and the other two scenarios ($p = 0.006$), as ~6\% fewer participants preferred physical sign in the smart meter scenario. 

\textbf{In-person notification is suggested in the Bluetooth beacon scenario and the Smart meter scenario.} Other than indirect notifications, participants in these two scenarios also suggested having their supervisor or people from upper management notify them in person about the data collection (n = 5). We did not find anyone suggesting this method in the Camera scenario. However, in the Camera scenario, one participant suggested implementing a written policy document.

\textbf{Across different means of notification, participants strongly preferred to be notified about the presence of a camera.} Participants' negative experiences with cameras could cause such preference. For example, one participant specifically mentioned their bad past experience with camera data collection as the reason why they wanted notification: ``I have been through abusive periods of my life revolving heavily around cameras (P50).''

\noindent\fbox{%
  \parbox{\linewidth}{%
    \textbf{Takeaway 5}: Email and physical signs were generally the most preferred means of notification for data collection in smart commercial buildings. However, one-size-fits-all should not be the strategy for notification. Fewer participants preferred physical signs for the smart meter scenario, which is significantly different from the other two scenarios. Therefore, a flexible selection of notification strategies (e.g., device-specific strategies) may be needed to inform occupants about the type and purpose of data collection by different IoT devices in smart commercial buildings.
  }%
}


\subsection{Potential Factors for Notification Preferences in Smart Commercial Building (RQ3)}

In this section, we discuss factors that may influence people's notification preferences for different IoT devices in smart buildings. Specifically, we focus on the following three factors: participants' awareness, confidence, and understanding of IoT devices in smart buildings based on participants' responses to the pre-scenario questions (i.e., general context). 

\subsubsection{Awareness of data collection}
We asked the participants whether they were aware of the data collection at their workplace. 52.44\% reported being aware, and 47.56\% reported being unaware. We further find that 91.5\% within the aware group and 89.74\% of participants within the unaware group reported wanting to be notified about the presence of data collection. This indicates that most occupants in smart commercial buildings may have concerns about their data being collected and thus want to be able to keep track of the data collection activities around them. It also confirms the importance of implementing notifications of data collection in commercial buildings to provide transparency.

Our result shows that the participants do not have significantly different notification preferences regardless of whether they are aware or unaware of data collection around them. However, regarding what people want to be notified about, we observed that about 4\% more participants in the aware group selected ``How your data can be used'', while for the other choices, there are slightly more participants (less than 4\%) in the unaware group. 
Regarding how people want to be notified, the response percentage for the physical sign is similar between the two groups. Mobile app (about 6\% more participants) and website (about 2\% more participants) are more preferred in the aware group, while email (about 7\% more participants) and paper document (about 3\% more participants) are more preferred in the unaware group. 

\subsubsection{Confidence with IoT}
We used a 5-point Likert scale question
to ask about participants' confidence levels with IoT technologies. For analysis purposes, we categorized ``1-Extremely unconfident'' and ``2-Somewhat unconfident'' into the unconfident group, ``3-Neither confident nor unconfident'' into the neutral group, and ``4-Somewhat confident'' and ``5-Extremely confident'' into the confident group. As a result, we had 16.3\% (80 out of 492) unconfident, 16.3\% (80 out of 492) neutral, and 67.5\% (332 out of 492) confident responses. We further found that the majority of participants within each group (96.3\% in unconfident, 92.5\% in neutral, and 88.9\% in confident) wanted to be notified about the presence of devices collecting data. This result suggests that even though people are confident with IoT technologies in general, they still prefer to be notified about the data collection around them.

Regarding what people want to be notified of, we found statistical significance ($p = 0.001$) for ``Presence of data collection'', ``How your data can be used'' ($p = 0.043$), ``For how long your data can be retained'' ($p = 0.000$), and ``Who can access your data'' ($p = 0.038$). For all of this data collection information, our pair-wise comparison further shows a significant difference between the confident group and the unconfident group ($p = 0.002$, $p = 0.039$, $p = 0.000$, respectively). 
Across all information about data collection, fewer participants in the confident group wanted to be notified as compared to the other two groups.

Regarding how people want to be notified, we did not find any statistical significance for the three groups. However, we observed that slightly more participants in the unconfident group preferred the other means of notification (i.e., physical sign, website, email, and paper document) rather than mobile app. There are about 12\% more participants in the confident group for mobile app selection than in the unconfident group. 

\subsubsection{Understanding of IoT}
We used a 5-point Likert scale question 
to ask the participants how they would describe their understanding of the IoT. For analysis purposes, we categorized ``1-No understanding'' and ``2-Below average'' into the below-average group, ``3-Average'' as the average group, and ``4-Above average'' and ``5-Strong understanding'' into the above-average group. Thus, we had 6.9\% (34 out of 492) below average, 35\% (172 out of 492) average, and 58.1\% (286 out of 492) above average. We further found that the majority of participants within each group (97\% in the below-average group, 90.1\% in the average, and 90.2\% in the above-average group) wanted to be notified about the presence of devices collecting data. This result shows that even though people claimed to have an average or above-average understanding of IoT technologies, they still want notifications about the data collection around them.

Regarding what people want to be notified about, we did not find any statistical significance for the three groups. However, across all information about data collection, we observed that there were slightly more participants in the below-average group who wanted to be notified. It is understandable that people with a below-average understanding of IoT may want to be notified of more information about data collection. 

Regarding how people want to be notified, we identified statistical significance for ``mobile app'' ($p = 0.009$) and ``email'' ($p = 0.021$). For both of these means of notification, our pair-wise comparison shows a significant difference between the average group and the above-average group ($p = 0.018$, $p = 0.049$, respectively). 
More participants in the above-average group preferred mobile app for notification, while more participants in the below-average group preferred email, physical sign, and paper document options.

\noindent\fbox{%
  \parbox{\linewidth}{%
    \textbf{Takeaway 6}: In smart commercial buildings, transparency of data collection is crucial. Even if people are aware of the data collection, are confident with IoT technologies, or are knowledgeable about IoT, they are still more likely to prefer receiving notifications about the data collection activities. The less confident people are with IoT technologies, the more information about data collection they want to be notified of. Regarding means of notification, people who claim to be confident with IoT technology and people who claim to have an above-average understanding of IoT tend to prefer mobile app over other means. This indicates the need for a flexible notification strategy that considers the background of the users.
  }%
}

\section{Discussion}
\label{sec:discussion}
In this section, we discuss the implications of privacy regulations and how our findings inform the design and operation of smart commercial buildings. We focus on how to inform and notify people about the type of data that is being collected, who may have access to their data, and how they can be better in charge of their own data. We further discuss the limitations of our study and potential future work.

\subsection{Policy Implications}

Recent privacy regulations around the world, including the European Union's General Data Protection Regulation (GDPR) and California Consumer Privacy Act (CCPA), lay the legal foundation for consumers to have control over the use and sharing of their personal information that businesses collect from them. Devices and systems that collect and use personal data about smart building occupants include: electronic access systems (such as smart entrances for exterior access points like gates and garages), thermostats, lighting, heating ventilation air conditioning (HVAC), sensors, voice recognition, and cameras. Businesses have to inform consumers about their data collection practices at or before the point of collection. Regardless of the method of collection, smart building owners must ensure visitors and guests are presented with clear and concise notice of the collection, use, and sharing of their personal information.

These laws lead to the fact that more respect and protection are being given to people's personal data, even at a heavy cost~\cite{huddleston2021privacy}. Therefore, to truly capitalize on the promise and benefits of smart buildings, it is not enough to minimize the data collected and limit the purpose of the data collection. Smart buildings need to take into account the occupants' \textit{privacy preferences and background} to determine a respectful way to \textit{notify} occupants of what data is being collected, how the collected data is utilized, and who has access to this data.

\subsection{Key Designs to Improve Smart Building Data Collection Transparency}
We detail our insights for designing data collection systems in smart commercial building settings as follows.

\textbf{Occupants in smart commercial buildings may be affected by legal ramifications as compared to occupants in private smart homes.} In particular, for smart homes, the owners can agree on installing IoT devices that collect their data and have control over them. However, this may not apply to people working in smart commercial buildings where there are much more complex interactions. It could lead to the fact that these occupants may not be aware of the data collection in the building where they work, and even if they are aware, they may have misunderstandings of the data collection activities.

As shown in our study results, we found that nearly half of the participants (47.56\%) were not aware of the collection of identifiable data about them. This result indicates that data collection in a smart commercial building is opaque to the occupants. The majority of participants (90.65\%) reported that they wanted to be notified about the presence of devices collecting data about them in their work environment. Thus, to improve the smart building occupant experience, notifications about data collection activities should be carefully considered when planning to deploy IoT devices.

Occupants in smart commercial buildings also have more concerns about their privacy as compared to those in smart homes: ``For starters, in my own home I can control what's going on regarding my own electronics. At my job I'm at the mercy of what the board/tech committee/whoever would make this decision, I think I at least deserve to know how my privacy is going to be violated. Also, all the HIPAA implications that would come with having all or at least a large part of conversations in the building being listened to at all times.'' Some were uncomfortable with being monitored at their workplace and that could affect their productivity: ``The data seems to be not related directly to my job, and I fear it could be used against me at some point. It feels like every aspect of my being is being monitored, and it does not create a comfortable environment. The camera sensor bothers me less than things like an energy sensor, because I know I am not doing anything that I would not like to be on camera, I am doing my job well. It's the random info that makes me uncomfortable, and I wonder just what is being done with it.''


\textbf{Notifications by email is consistently the most preferred method for the participant's preference on how to be notified across scenario and confidence groups.} On the contrary, notification by mobile app is one of the lowest rated methods. This result shows a very different notification preference as compared to a smart home environment where people preferred mobile app notification the most~\cite{thakkar2022would}. Although notification by email is preferred in the smart commercial building context, it is important to note that email notification is only practical in scenarios where all data subjects can have their email addresses registered for notifications.

We suspect that the amount of user effort involved in receiving notifications is one of the deciding factors in the smart commercial building context. For example, emails require the least amount of user effort in the professional context, given that email is the most common way of communication at work. Similarly, physical signs, websites, or paper documents likely will not require the users to actively do anything to subscribe to the notifications. However, to receive notifications through a mobile app, users need to rely on an additional device, and installing an app also takes some effort.

\textbf{In terms of what to notify the occupants, people seem to be less interested in: 1) data collected by smart meters, and 2) the duration with which data is retained.} These results indicate that if a designer is looking to reduce fatigue by minimizing the amount of information in the notification, information about sensors similar to smart meters and duration information can be the first candidates to eliminate. When looking at the types of users, we found that factors such as confidence about IoT and IoT knowledge impact users' privacy notification preferences in different data collection scenarios. 
These results indicate that a notification system can likely take advantage of a user's self-reported understanding and confidence of IoTs to tailor the notification rate for privacy updates and changes. 

\textbf{People consider certain locations in the building (e.g., bathrooms) as private spaces where data should not be collected.} Some participants mentioned that they do not want to be recognized while using the bathroom as there is no privacy. Even in the smart meter scenario, participants were okay with water use tracking but were hesitant about collecting data in private areas. Some participants in this scenario mentioned that water usage is not so much of a big deal, but they do not like the idea of someone keeping track of their bathroom usage. They further explained that it would be embarrassing if a supervisor or manager came by and questioned their usage. These results might indicate that smart building managers should consider setting different policies for data collection in public, semi-public, and private spaces in smart buildings.


\textbf{It is important for smart buildings to create a fiduciary relationship with their occupants by aligning interests.}
Our findings (Section~\ref{result:dataaccess}) indicate that many occupants feel the IoT data collection in smart buildings benefits others (e.g., building managers, employers) more than themselves, which may lead to common negative perceptions or misconceptions around unwanted surveillance. Although some of our participants recognized the potential personal benefits of such IoT data collection, it is not currently feasible for them to take full advantage of the data being collected. To improve the acceptance and resolve potential privacy concerns, it is critical for smart building managers to create a fiduciary relationship with occupants by aligning the interests of both parties. 

We recommend that smart buildings should ensure occupants are aware of their personal benefits from various IoT data collection, which could be conveyed through effective privacy notifications in occupants' preferred formats. Also, necessary software infrastructure (e.g., APIs) that enables occupants to access and take advantage of certain collected IoT data for their personal benefits will contribute to building such a fiduciary relationship.

\subsection{Limitations and Future Work}
Our study has some limitations. First, our results rely on participants' self-reported data. 
To mitigate the bias, we cross-checked participants' answers throughout the survey to ensure their responses were consistent, indicating a satisfactory level of trustworthiness regarding their preferences. Second, we used hypothetical scenarios in our study to prompt participants' preferences, and the preferences can vary based on context. Third, some of our participants may not have experience with smart buildings based on our definition. However, it is worth noting that in the survey, we asked whether participants had seen any smart devices in their building. The results indicated that the majority of our participants came across some IoT devices at their workplaces. Lastly, some participants may be biased due to the same questions regarding their preferences before and after we presented the scenarios. 

Our study in this paper focuses on smart commercial building occupants in the US. However, as more countries and regions deploy smart building technology in real-world settings, future studies can consider cross-cultural perspectives. For example, our finding regarding occupants' perceptions of data access can be further extended to identify the differences based on different work cultures and regulations. Additionally, future work can potentially do a field study to further explore people's notification preferences in real smart building settings that consider different contexts. It is also interesting to explore workplace monitoring or tension between employees and employers. Some small-scale interviews or case studies at different types of smart building workplaces would be ideal to collect such data.

Our study is the necessary step toward designing a more effective data collection and disclosure scheme for smart buildings. Future research can look into building personalized notification systems for data collection in smart buildings. However, a complex multi-stakeholder environment could be a big challenge to deploying such personalized systems. Thus, exploring the complex relationships and potential conflicts between different stakeholders in the smart building context is also important.


\section{Conclusion}
\label{sec:conclusion}
Smart buildings and their applications are becoming more popular in urban areas around the world. They increasingly adopt IoT technologies to manage their resources and services. Large deployments of interconnected sensors, actuators, and smart devices in smart buildings improve productivity and user experience across application domains. However, this means pervasive, multi-modal, continuous, and scalable data collection in such a semi-public space. This is also an underexplored domain. Therefore, our goal in this study is to understand the occupants' perceptions of data collection and their notification preferences for different data collection scenarios in smart commercial buildings. We conduct a user study with 492 participants who are occupants of smart commercial buildings. Our analysis results show that many participants (47.56\%) are unaware of the data collection, and email is, in general, the most preferred means of notification. We also find that people have different preferences for notifications when they face different data collection scenarios. Our findings will help guide the design and implementation of privacy-related notifications in smart buildings and increase occupants' awareness of the nearby data collection. In the bigger picture, future smart cities can use the insights from this research to develop privacy-respecting infrastructure and effective notification schemes.
\section*{Acknowledgements}
This research was supported in part by the National Science Foundation (NSF) Secure and Trustworthy Computing program (Grants CNS-1801316, CNS-2320903). Other grants supporting this research included NSF CNS 2114074, NSF OAC 2002985, NSF CNS 1943100, NSF OAC 1920462, NSF CNS 2323105, and Google Research Scholar Award. The US Government is authorized to reproduce and distribute reprints for Governmental purposes, notwithstanding any copyright notices thereon. The views and conclusions contained herein are those of the authors and should not be interpreted as representing the official policies or endorsements, either expressed or implied of the US Government or other funding agencies. This research was also partially funded by an internal grant at the University of Maryland, Baltimore County. We thank Clay Ford from UVA Stat Lab for providing statistical consulting and UVA Link Lab for providing feedback on the study design. We also thank our shepherd and anonymous reviewers for their constructive comments on this research.

\bibliographystyle{plain}
\bibliography{references}

\appendices

\section{Background Details of Participants}
\label{appendix:participants}
\begin{figure}[htbp]
  \centering
\includegraphics[width=0.5\textwidth]{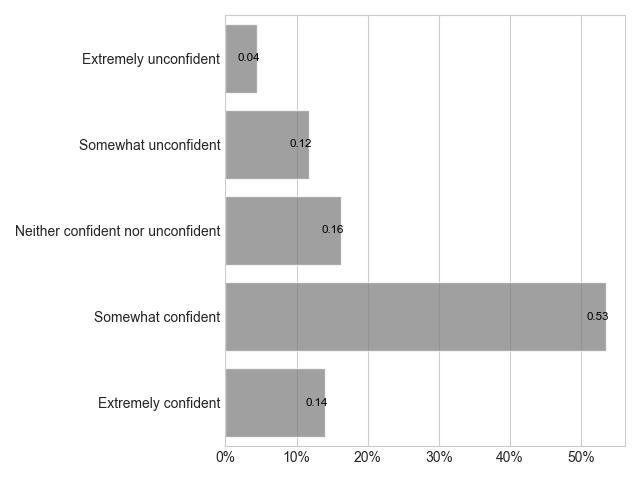}
  \caption{Participants' level of confidence with the IoT technologies on a 5-likert scale (1-Extremely unconfident to 5-Extremely confident).}
  \label{fig:confidenceiot}
\end{figure}

\begin{figure}[htbp]
  \centering
    \includegraphics[width=0.5\textwidth]{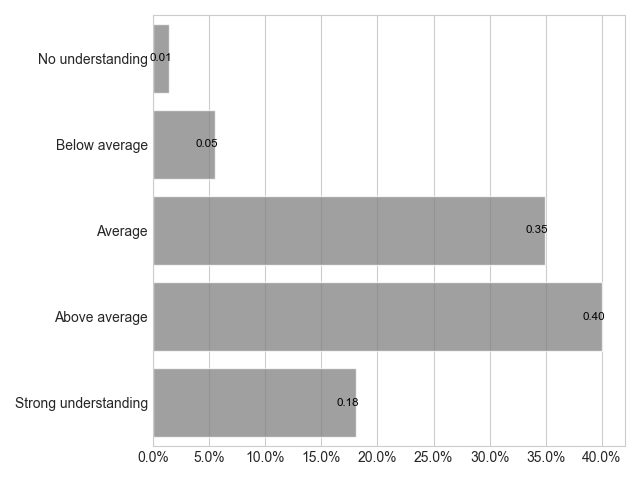}
  \caption{Participants' understanding of the IoT technologies on a 5-likert scale (1-No understanding to 5-Strong understanding).}
  \label{fig:understandingiot}
\end{figure}

In Figure~\ref{fig:confidenceiot} and Figure~\ref{fig:understandingiot}, we show participants' level of confidence with the IoT technologies and participants' understanding of the IoT technologies, respectively. 

\section{Survey Instrument}
\label{appendix:surveyinstrument}

\subsection{Screening}

\Qitem{ \Qq{Which of the following working environments best describe where you have worked in or are currently working in? (Choose all that apply)}
\begin{Qlist}
\item I work from home
\item I have a private office to myself
\item I share a private office with some colleagues
\item I work in an open workspace with a designated desk
\item I work in an open workspace without a designated desk
\item I work outdoors
\item I work in a retail/service sector (e.g., restaurants, retail stores, grocery stores)
\item Other: \Qline{4cm}
\end{Qlist}
}

\subsection{Main Survey}

\centerline{--- Background and Awareness Questions ---}

\Qitem{ \Qq{What devices do you notice at your workplace? (Choose all that apply)}
\begin{Qlist}
\item Camera
\item Motion sensor
\item Energy sensor
\item Water sensor
\item Light sensor
\item Temperature sensor
\item Smart TV
\item Smart speaker
\item Other: \Qline{4cm}
\item None
\end{Qlist}
}

\Qitem{ \Qq{Are you aware of the devices' data collections at your workplace?}
\begin{Qlist}
\item Yes
\item No
\end{Qlist}
}

\centerline{--- Perception of Data Collection Questions ---}
\begin{itemize}
    \item Scenario 1 (Bluetooth beacons): Suppose your employer installs Bluetooth beacons (devices that wirelessly broadcast a unique identifier to nearby electronic devices) at your workplace. These beacons are used to collect location and movement information to understand how the space is being used.)
    \item Scenario 2 (Cameras): Suppose your employer installs video surveillance cameras at your workplace that collect photo/video footage to ensure workplace security.
    \item Scenario 3 (Smart meters): Suppose your employer installs smart meters at your workplace that collect data about human activities and resource usage (e.g., energy consumption, bathroom usage) to monitor and optimize the building's resource consumption.
\end{itemize}

\Qitem{ \Qq{In the scenario, who do you think will have access to the collected data about you? (Choose all that apply)}
\begin{Qlist}
\item My building manager
\item My supervisor
\item The government
\item The manufacturer of the devices
\item My company
\item Myself
\item Other: \Qline{4cm}
\end{Qlist}
}

\Qitem{ \Qq{In the scenario, who are you comfortable with having access to the collected data about you? (Choose all that apply)}
\begin{Qlist}
\item My building manager
\item My supervisor
\item The government
\item The manufacturer of the devices
\item My company
\item Myself
\item Other: \Qline{4cm}
\end{Qlist}
}

\Qitem{ \Qq{In the scenario, who do you think will benefit from having access to the collected data about you? (Choose all that apply)}
\begin{Qlist}
\item My building manager
\item My supervisor
\item The government
\item The manufacturer of the devices
\item My company
\item Myself
\item Other: \Qline{4cm}
\end{Qlist}
}

\Qitem{ \Qq{In the scenario, which of the following purposes do you think the collected data about you might be used for? (Choose all that apply)}
\begin{Qlist}
\item User profiling
\item Localization
\item Enforcing policies
\item Improving smart devices
\item Other: \Qline{4cm}
\item None
\end{Qlist}
}

\centerline{--- Notification Preferences Questions ---}

Note that for post-scenario, we replace ``at your workplace'' with ``in the scenario'' to apply the context.

\Qitem{ \Qq{Do you want to be notified about the presence of the devices collecting data about you at your workplace?}
\begin{Qlist}
\item Yes
\item No
\end{Qlist}
}

\Qitem{ \Qq{Please briefly explain why you want (or why you don't want) to be notified:}

\Qline{8cm}
}

\Qitem{ \Qq{At your workplace, which of the following do you want to be notified about? (Choose all that apply)}
\begin{Qlist}
\item Presence of data collection (including types of data being collected)
\item Purposes for which your data can be used
\item How your data can be used
\item For how long your data can be retained
\item Who can access your data
\item None of the above
\end{Qlist}
}

\Qitem{ \Qq{How do you want to be notified? (Choose all that apply)}
\begin{Qlist}
\item Physical sign
\item Mobile app
\item Website
\item Email
\item Paper document
\item Other: \Qline{4cm}
\end{Qlist}
}

\centerline{--- Demographic Information ---}

\Qitem{ \Qq{With which gender identity do you most identify?}
\begin{Qlist}
\item Male
\item Female
\item Other: \Qline{4cm}
\item Prefer not to answer
\end{Qlist}
}

\Qitem{ \Qq{What is your age group?}
\begin{Qlist}
\item 18 - 24 years old
\item 25 - 34 years old
\item 35 - 44 years old
\item 45 - 54 years old
\item 55 - 64 years old
\item 65 - 74 years old
\item 75 years or older
\item Prefer not to answer
\end{Qlist}
}

\Qitem{ \Qq{What is the highest level of education you have completed?}
\begin{Qlist}
\item Some high school
\item High school graduate
\item Some college
\item Associate's degree (2-year college)
\item Bachelor's degree (4-year college)
\item Graduate degree (Masters, PhD, MD, JD, etc.)
\item Other: \Qline{4cm}
\item Prefer not to answer
\end{Qlist}
}

\Qitem{ \Qq{What is your annual personal income before taxes (USD)?}
\begin{Qlist}
\item Less than \$10,000
\item \$10,000 - \$24,999
\item \$25,000 - \$49,999
\item \$50,000 - \$74,999
\item \$75,000 - \$99,999
\item \$100,000 - \$149,999
\item \$150,000 and greater
\item Prefer not to answer
\end{Qlist}
}

\centerline{--- Confidence and Understanding ---}

\Qitem{ \Qq{How confident are you with the Internet of Things technologies in general (e.g., voice assistants, smart security camera, drones, smart watches, virtual reality glass, etc.)?}
\begin{Qlist}
\item Extremely unconfident
\item Somewhat unconfident
\item Neither confident nor unconfident
\item Somewhat confident
\item Extremely confident
\end{Qlist}
}

\Qitem{ \Qq{How would you describe your understanding of the Internet of Things?}
\begin{Qlist}
\item No understanding
\item Below average
\item Average
\item Above average
\item Strong understanding
\end{Qlist}
}

\end{document}